# Self-Diffusion and Structure of a Quasi Two-Dimensional, Classical Coulomb Gas Under Increasing Magnetic Field and Temperature

J. D. Hernández Velázquez[1], Z. Nussinov[2, 3], and A. Gama Goicochea[1*]

[1]Tecnológico Nacional de México, Tecnológico de Estudios Superiores de Ecatepec, División de Ingeniería Química y Bioquímica, Ecatepec de Morelos, Estado de México 55210, Mexico

[2]Department of Physics, Washington University, St. Louis, Missouri 63130, USA

[3]Rudolf Peierls Centre for Theoretical Physics, University of Oxford, Oxford 0X1 3PU, United Kingdom

## Abstract

The influence of a magnetic field applied perpendicularly to the plane of a quasi-two-dimensional, low density classical Coulomb gas, with interparticle potential $U(r) \sim 1/r$, is studied using momentum-conserving dissipative particle dynamics simulations. The self-diffusion and structure of the gas are studied as functions of temperature and strength of the magnetic field. It is found that the gas undergoes a topological phase transition when the temperature is varied with, in accord with the Bohr-van Leeuwen (BvL) theorem, the structural properties being unaffected, resembling those of the strictly two-dimensional Kosterlitz-Thouless transition, with $U(r) \sim \ln(r)$. Consistent with the BvL theorem, the transition temperature and the melting process of the condensed phase are unchanged by the field. Conversely, the self-diffusion coefficient of the gas is strongly reduced by the magnetic field. At the largest values of the cyclotron frequency, the self-diffusion coefficient is inversely proportional to the applied magnetic field. The implications of these results are discussed.

---

[*] Corresponding author. Email: agama@alumni.stanford.edu

# I INTRODUCTION

The globally neutral, two-dimensional (2d) Coulomb gas is a system with equal number of positively and negatively charged disks whose electrostatic interaction is given by $U(r) \sim q^2 \ln(r)$, with $q$ being the charge on the disks, as follows from solving Poisson's equation in 2d. Kosterlitz and Thouless (KT) showed that a topological phase transition takes place in this type of system at temperature $T_C = q^2/4$ [1]. Above $T_C$ the gas is conducting while below $T_C$, is a dielectric. This transition is also characterized by the lack of a well-defined order parameter and no symmetry is broken at the phase transition. A peak in the specific heat appears at a temperature slightly above $T_C$, and the spatial correlation function of the disks decays algebraically below $T_C$ with a temperature dependent exponent $\eta(T)$ such that $\eta(T_C) = 1/4$ [1]. Above $T_C$ the correlation function decays exponentially. This is the basic phenomenology of the KT transition [2,3].

In this contribution we focus on studying the effects of applying a magnetic field, perpendicularly to the plane of a quasi-2d Coulomb gas having electrostatic interactions of the $U(r) \sim 1/r$ type, as the temperature and strength of the magnetic field are varied. Recent work [4] demonstrates that an overall neutral system of charged spheres under quasi-2d confinement undergoes a topological phase transition similar to that of its strictly-2d counterpart. In particular, the critical temperature is given by the KT prediction divided by a constant length scale, the correlation functions decay algebraically with distance below $T_C$, albeit with an exponent larger than $1/4$ at $T_C$. Above $T_C$, the correlations decay exponentially. There is a maximum in the specific heat close to $T_C$; the translational order parameter is size dependent and disappears in the thermodynamic limit. As the thickness of the quasi-2d Coulomb gas increases and the system becomes fully three-dimensional, the topological phase transition disappears [4].

The importance of these types of studies lies in the need to better understand low-dimensional (quasi-2d) systems, since there are no truly two-dimensional materials. Recent examples are quasi-2d electron gases (q2DEG) found at the interfaces of rare-earth oxide heterostructures such as LaTiO3-SrTiO3 [5,6], LaAlO3-SrTiO3 [7–9] and LaAlO3-EuTiO3-SrTiO3 [10]. They have generated interest in the last 20 years because of their remarkable electronic and magnetic properties. In some experiments, the q2DEG can be confined within a layer up to 2 nm thick [5]. On the other hand, there exists experimental evidence showing that the magnetoresistance of the q2DEG in oxide heterointerfaces can be modulated with the orientation of the external magnetic field [11,12], displaying rather different behavior from its three-dimensional counterpart [11]. The purpose of this work is to determine to what extent the properties of such a topological phase transition are modified when a magnetic field is applied. This problem is important not only because of its relevance in energy storage, materials design, and device engineering [13–15], but also because several aspects at the basic science level remain poorly understood. Firstly, the question arises as to what extent are the structural properties of a quasi-2d Coulomb gas affected by the presence of a perpendicular magnetic field. It is also of crucial importance to determine how the thermodynamic and transport properties of this type of system are modified by a transverse magnetic field, as it happens with more ordinary phase transitions, such as in superconductors. Also, this knowledge can potentially increase the number of applications. The Bohr-van Leeuwen (BvL) theorem, applicable to general classical systems, states that the free energy of the equilibrium system is invariant under the application of an external magnetic field. The well-proof of the BvL theorem by a trivial shift of momentum space integration variables in the partition function integral [16] can be replicated when the Hamiltonian is augmented by external fields. This invariance also holds when additional external fields of infinitesimal strength are added to the Hamiltonian. The derivatives of the free energy relative to such additional fields yield general correlation functions which, by the BvL theorem, must similarly be invariant under the application of an external magnetic field. This includes, in particular, also correlation functions that provide structural measures. Thus, a trivial generalization of the BvL theorem implies that general equilibrium correlation functions including correlation functions associated with structural measures must, rigorously, remain invariant when a magnetic field is present. This conclusion relates to thermodynamic and such general time independent equilibrium correlations structural properties. The free energy cannot provide information on time dependent correlation functions such as the velocity autocorrelation function (whose integral yields, by the Green-Kubo relation, the self-diffusion constant). Thus, while the long time average equilibrium properties are invariant (by the BvL theorem and the trivial generalization thereof discussed above) under the application of the magnetic field, no such general statements follow for general time dependent quantities. Studying the effect of the magnetic field on the system dynamics is a central objective of our work. In what follows, we first illustrate that our numerical results for long time averaged static equilibrium quantities remain invariant under the application of the magnetic field as required by the BvL theorem. Following this verification of our *in silico* results, we then proceed to explicitly study, for the first time, the effects of an applied external magnetic field on the dynamical properties (principally, the self-diffusion coefficient) of these quasi-2d systems.

There are only a few reports concerning the low density, 2d-Coulomb gas under a transverse magnetic field. One of first is the work of Hansen and collaborators [17], who calculated the self-diffusion coefficient of a system of negatively charged disks in a positive background. The self-diffusion coefficient $D$ was obtained for several values of the coupling constant, Γ (defined as the ratio of the electrostatic energy to the thermal energy), finding that smaller values of Γ yielded larger values of $D$. They concluded that the magnetic field did not affect the self-diffusion coefficient of the charged disks [17]. This result is not expected, since the circular motion of the charges produced by the magnetic field precludes diffusion. Yamada and Ferry [18] found that structural properties, such as the radial distribution function, displayed no change when the intensity of the magnetic field was increased. Dubey and Gumbs [19] modeled a low density set of negatively charged particles in a positive background in 2d, using molecular dynamics. Neither the structure nor the thermodynamics of their system were qualitatively affected by magnetic fields, though the self-diffusion coefficient was not calculated. A follow-up publication [20] studied a 2d system of negative charges in a positive background under a 2d modulating potential in the *xy*-plane subject to a perpendicular magnetic field. In the absence of the latter potential, $D$ decreased monotonically with magnetic field; both cases were studied at fixed temperature [20]. Other studies have modeled 2d charges under a perpendicular magnetic field, with the electrostatic interaction given by $U(r) \sim 1/r$, such as that of Ranganathan and coworkers [21]. Their self-diffusion coefficient decreases monotonically when the magnetic field is increased up to 5 T [21]. Feng and collaborators [22] find that the self-diffusion coefficient decays as the transverse magnetic field increases, as $D \sim 1/(1 + c\beta^2)$. Here, $\beta = \omega_C/\omega_P$ is the ratio of the cyclotron frequency over the plasma frequency and $c$ is a constant [22]. Similar systems are modeled by Ott *et al.* [23], whose self-diffusion coefficient is seen to decays as $D \sim 1/\beta$ under strong transversal magnetic field [23], in contradiction to Feng *et al.*'s predictions [22]. In most of these works, the authors modeled charged disks in strictly 2d, with logarithmic electrostatic interactions. However, to compare with experiments and to predict properties that are useful for applications of quasi-2d systems, one should model spheres instead of disks, since no material is truly two dimensional. That is the focus of this work.

## II MODELS, METHODS, AND SIMULATION DETAILS

The system studied here is a low density, quasi – 2d Coulomb gas of spheres confined to move in a box whose thickness is small but finite, by means of numerical simulations. The motion of the particles is obtained from the solution of the dissipative particle dynamics (DPD) force field [24,25], complemented with the 3d-Ewald sums for confined systems, to account for the long-range nature of the electrostatic interactions [26]. The charged spheres are restricted to move under quasi-2d confinement by effective, short-range wall forces acting only along the *z*-axis, which is the confining direction [27]. All quantities are reported in reduced units, unless stated otherwise, and are represented by asterisked symbols. The number density of the gas is set to $\rho^* = 0.03$ for all cases reported here. The volume of the simulation box is $L_x^* \times L_y^* \times L_z^* = 80 \times 80 \times 1 r_C^{*3}$, where $r_C^* = 1$ is the length scale in DPD [28]. In all cases, the charge on the particles is fixed to $|q^*| = 4$ and the total number of particles is $N = 200$, unless stated otherwise. Lengths are reduced by the DPD forces' cutoff radius, $r_C = 6.46$ Å [28], so that $L^* = L/r_C$, and similarly for all other units [29]. All simulations are run for at least $10^7$ time steps. The magnetic field is varied from $B_Z^* = 0$ up to $B_Z^* = 0.1$ [29]. The contribution of the perpendicular magnetic field to the dynamics of the system is introduced through the Lorentz force. Full details have been provided elsewhere [4] and are therefore omitted here for brevity. All additional details pertinent to this work can be found in the Supplemental Material [29].

## III RESULTS AND DISCUSSION

As we briefly reviewed in the Introduction, by the BvL theorem and a trivial extension thereof, the thermodynamic and structural properties of our system must remain invariant under the application of an external magnetic field. In what follows, we begin by demonstrating that our numerical results indicate that thermodynamic and structural measures are indeed independent of the applied field. Armed with these results, we then turn to our main objective of studying the system dynamics and illustrate that the diffusion constant is strongly dependent on the applied field. To track changes in the structure of the quasi-2d Coulomb gas when the strength of the transverse magnetic field and the temperature are increased, the translational order parameter (TOP), is calculated as follows [30]:

$$\Psi_T = \frac{1}{N} \langle \left| \sum_{j=1}^{N} e^{i\vec{K}\cdot\vec{r^*}_j} \right| \rangle. \tag{1}$$

In eq. (1), $N$ is the total number of charged particles, $\vec{K}$ is the first shell reciprocal lattice vector, and the angular brackets represent an average over time. To determine $T_C^*$, the TOP is calculated for temperatures in the range $0.375 \leq T^*/T_C^* \leq 11.14$, defining $T_C^*$ as the temperature at which the TOP has the steepest change [4]. The $T_C^*$ obtained with this approach is found to be in excellent agreement with the KT prediction, $T_C^* = q^{*2}/4r_C^*$ [1]. This procedure is conducted for transverse magnetic field in the range $0.00 \leq B_Z^* \leq 0.10$. The results are presented in Fig. 1, showing that the TOP is close to one below $T_C^*$, where the charges are condensed into a single structure, independently of the strength of the magnetic field. The temperatures are normalized by the value of $T_C^*$, found as previously described. Above $T_C^*$, most of the charges are unpaired and the TOP is close to zero. The magnetic field does not change the critical temperature, $T_C^*$, of a low density, quasi-2d Coulomb gas. This is one of our main conclusions.

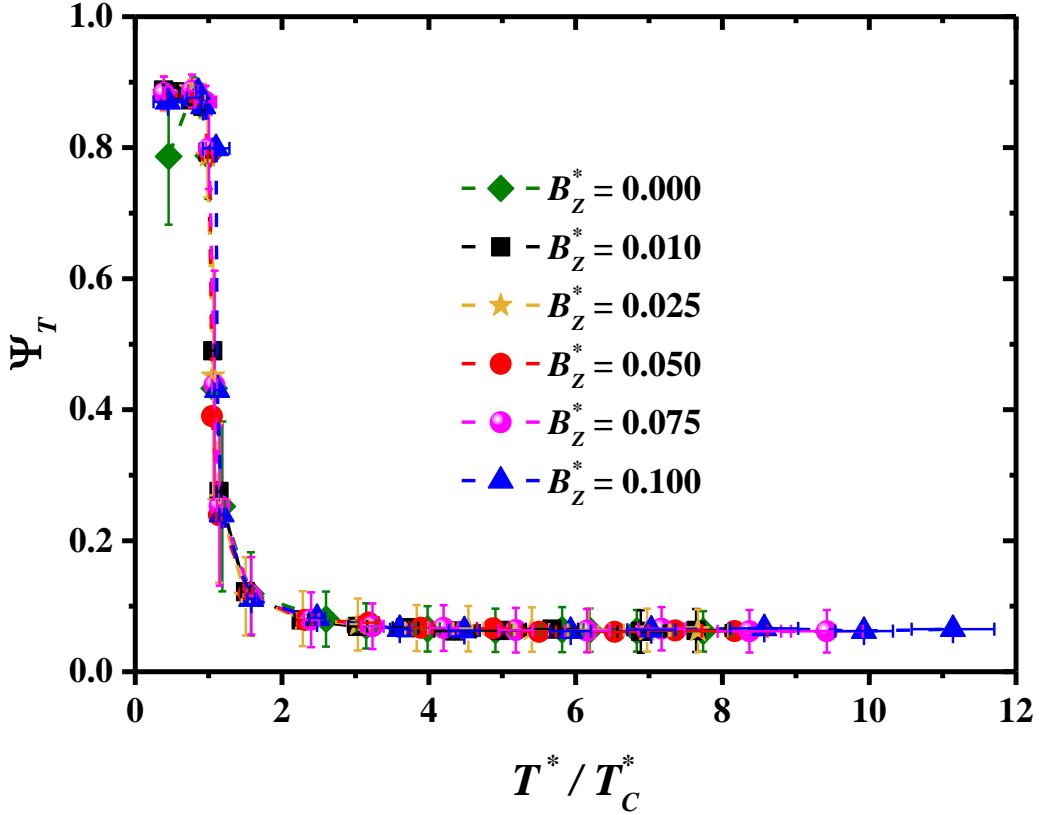

**Figure 1.** The translational order parameter, $\Psi_T$, as a function of the normalized temperature, $T^*/T_C^*$ for increasing values of the external magnetic field, $B_z^*$, applied perpendicularly to the *xy*-plane of the slit on which the charged spheres move. The dashed lines are guides for the eye. The data for $B_Z^* = 0.0$ were taken from [4].

To study the structure and the spatial correlations of the quasi-2d Coulomb gas, the radial distribution function (RDF), $g(r)$, is calculated as the temperature is increased, for particles of opposite charge, under the influence of a constant magnetic field. Figure 2 shows the RDF at a temperature (a) below $T_C^*$ and (b) above $T_C^*$. For simplicity, only results for the minimum and maximum values of $B_z^*$ are shown. For charged spheres with no external field, spatial correlations decay algebraically with relative distance at $T^* < T_C^*$, in agreement with predictions for strictly-2d charges [1,4]. This is afforded in Fig. 2(a) by a comparison between the raw data (black curve) and the power law fit (dashed red line). Consistent with the BvL theorem, the application of the maximum transverse field used in this work introduces (up to negligible numerical error) somewhat sharper oscillations. However, it does not change this algebraic-decay behavior, as displayed in Fig. 2(a) by the blue line.

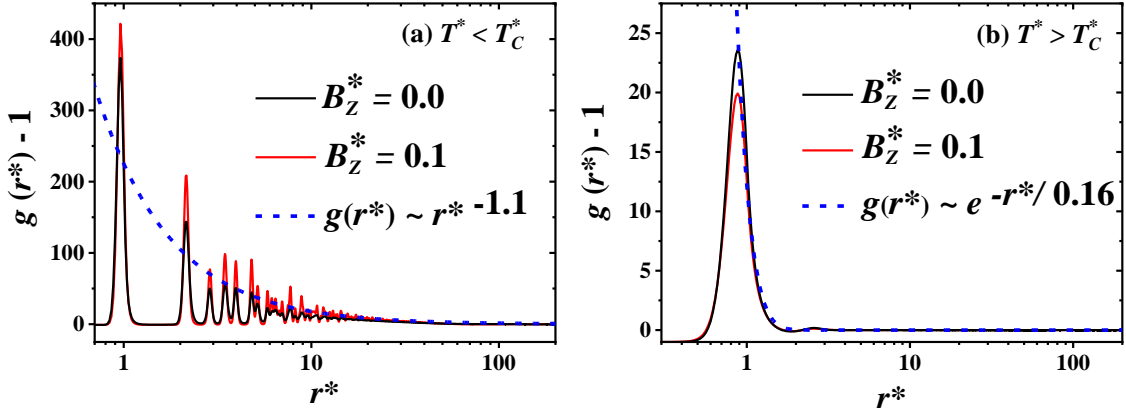

**Figure 2.** Radial distribution functions, $g(r^*) - 1$, between charges of opposite sign at (a) $T^* = 0.75 T_C^*$, and at (b) $T^* = 7.5 T_C^*$, as functions of relative distance, $r^*$, in reduced units. Only data for the maximum and minimum strengths of the applied magnetic field are shown, for simplicity. The data are plotted in semi-log scale, rather than log-log scale, to avoid having broken lines along the $y$-axis when $g(r^*) - 1$ becomes negative. The dashed blue lines in (a) and (b) are best fits to algebraic and exponential decays, respectively. In all cases, there are $N = 3 \times 10^4$ charged spheres, with number density $\rho^* = 0.03$ and charge $|q^*| = 4.0$. The data for $B_Z^* = 0.0$ were taken from [4].

Above the critical temperature of this topological phase transition, the spatial correlations decay exponentially [4], as in the 2d KT-transition [1]; see Fig. 2(b). At $T > T_C^*$, the charged spheres dissociate and the gas becomes conducting. The application of a transverse magnetic field does not change the structure of the quasi-2d Coulomb gas above its $T_C^*$, as shown by the blue line in Fig. 2(b), just as it did not change it below $T_C^*$. This relative insensitivity of the RDF to the applied magnetic field agrees with previous findings [19,20], using molecular dynamics simulations for a strictly-2d Coulomb gas under a perpendicular magnetic field.

The melting of the low-temperature condensed phase is also unaffected by the application of a transverse magnetic field, as shown in Fig. 3. Here $n_F$ represents the number density of free charges, namely, those that are not forming dipole pairs. The solid red line in Fig.

3 is an approximate analytical solution, $n_F = \rho^*(r_0/\lambda)^{1/2(T/T_C)}$, found by Minnhagen [3], for the strictly-2d Coulomb gas. Here, $\rho^*$ is the number density, $r_0$ is maximum extension of the charge distribution and $\lambda$ is the screening length. As the temperature is raised, progressively more dipole pairs disintegrate yet the magnetic field does not modify the density of free particles, as seen in Fig. 3. The melting process of the condensed phase is not affected by the magnetic field. This is another one of our main conclusions, which is supported by experiments [31], as well as by the data shown in Fig. 1.

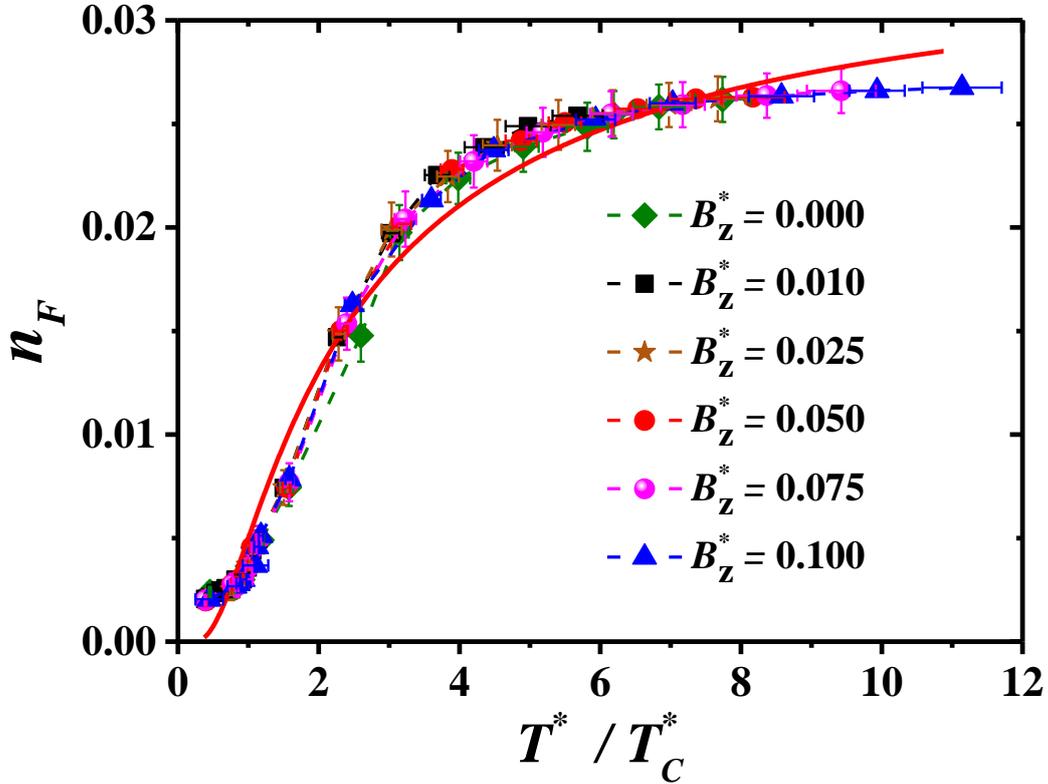

**Figure 3.** The melting of the system, measured by the proliferation of the number density of free charges, $n_F$, as the temperature is raised and for increasing strength of the applied magnetic field. The data for $B_Z^* = 0.0$ were taken from [4]. The solid red line is an approximate analytical expression for strictly-2d disks [3], $n_F = \rho^*(r_0/\lambda)^{1/2(T/T_C)}$, where $\rho^* = 0.03$ is the absolute number density, $r_0$ is maximal extension of the charge distribution and $\lambda$ is the charge screening length, with $(r_0/\lambda) = 0.6$. See text for details.

We now turn to our central objective of studying the system dynamics. Starting from the calculation of the mean square displacement of the charged particles, their self-diffusion coefficient, $D^*$, is obtained [29]. At temperatures $T^* \leq T_C^*$, the data show that $D^*$ is almost negligible for all values of the applied magnetic field; see Fig. 4. The magnetic field strongly reduces the self-diffusion coefficient, specially at low temperatures, as seen in Fig. 4. At $T^*/T_C^* < 1$ the diffusion is negligible, which is expected because at those temperatures the system is condensed, see Fig. S1(a) in [29], hence the Lorentz force is zero. For $T^*/T_C^* > 1$ there is diffusion that increases with increasing temperature, but it is strongly suppressed by the magnetic field. The growing cyclotron frequency favors circular motion of the charges over their diffusion. A large increase in the self-diffusion coefficient is found at $T^* = 6T_C^*$ for $B_Z^* = 0$ in Fig. 4. This is a consequence of the fact that, around that temperature, the number of free (unbound) charges per unit volume has almost reached the value of the global number density, see Fig. 3. Since most charges are unpaired, the temperature is high and there is no magnetic field to induce circular motion, their diffusion is maximized.

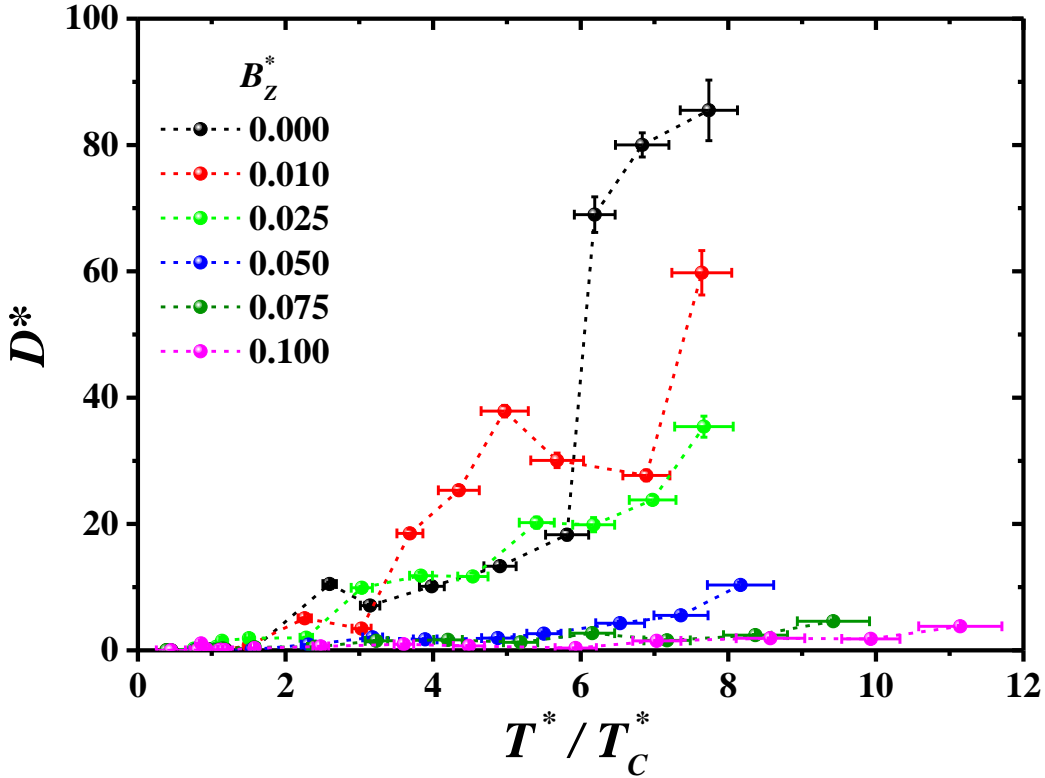

**Figure 4.** The self-diffusion coefficient of the charged spheres, $D^*$, as a function of the reduced temperature, $T^*/T_C^*$, for increasing strength of the transverse magnetic field, $B_Z^*$. The dashed lines are only guides for the eye.

To compare our results with previously published work on dusty plasmas in strictly 2d [22,23,32], we plot $D^*$, normalized by its value for $B_Z^* = 0$ at $T^*/T_C^* = 7.5$, called $D_0^*$, as a function of $\beta^* = \omega_C^*/\omega_P^*$, in Fig. 5. The self-diffusion coefficient $D^*$ is reduced when increasing $\Gamma^*$ and $\beta^*$, with $\Gamma^* = q^{*2}/a^*T^*$ being the coupling constant, and $a^* = (4\pi\rho^*/3)^{-1/3}$ is the Wigner-Seitz radius. The results follow the same trend seen in experiments on dusty plasmas under strong magnetic field [31]. The solid lines in the main panel in Fig. 5 are best fits to $D^* \sim 1/\beta^*$, which is approximately fulfilled for all four $\Gamma^*$ values. This so-called Bohm diffusion type [33] has also been observed in one component and binary charged systems in strictly 2d with Yukawa interactions [23]. A strong drop in the self-diffusion coefficient with increasing magnetic field has been observed in molecular dynamics simulations of 3d plasmas by Vidal and Baalrud [34]. They report that data on the component of the self-diffusion coefficient perpendicular to the magnetic field ($D_\perp$), which also follows approximately Bohm's diffusion for relatively high magnetic fields ($\beta \gtrsim 0.1$) [34]; see the inset in Fig. 5.

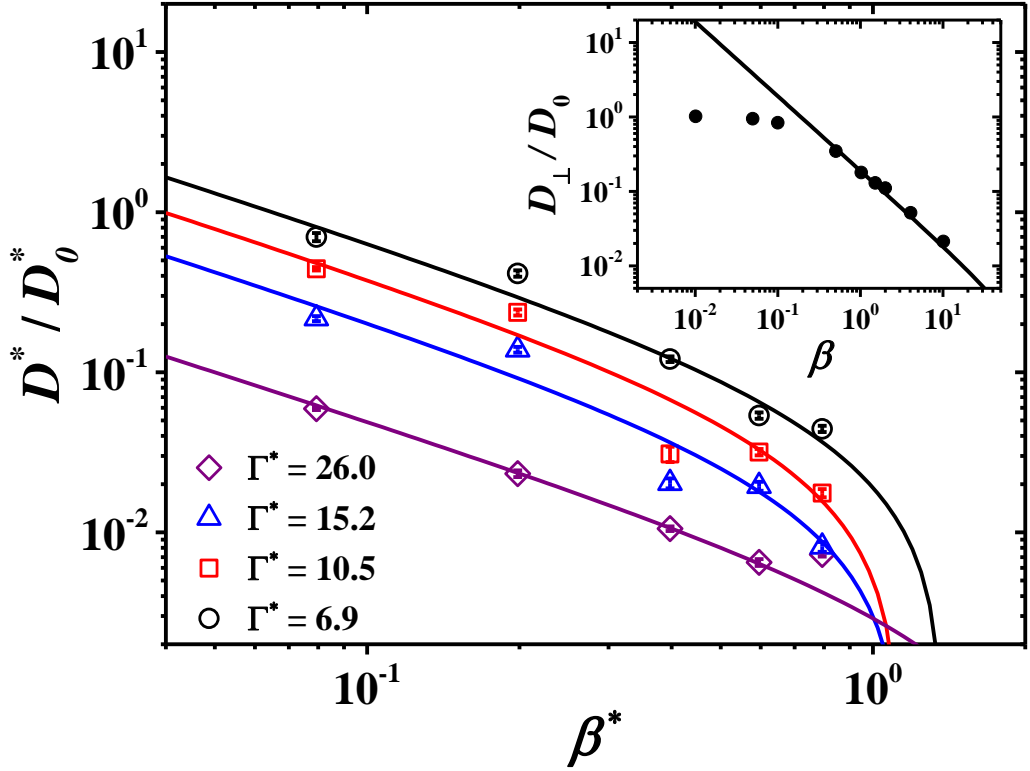

**Figure 5.** Self-diffusion coefficient, $D^*$, normalized by its value without magnetic field at $T^*/T_C^* = 7.5$, $D_0^*$, as a function of the parameter $\beta^* = \omega_C^*/\omega_P^*$, for four values of the coupling constant, $\Gamma^*$, for $T^*/T_C^* > 1$ in all cases. Here, $\omega_C^*$ and $\omega_P^*$ are the cyclotron and plasma frequencies, respectively. The inset shows the normalized self-diffusion coefficient, perpendicular to the magnetic field ($D_\perp/D_0$) reported by Vidal and Baalrud [34], for a system of $N = 5 \times 10^3$ particles and $\Gamma = 10$. The solid lines are the best fits to the function $D^*/D_0^* = \alpha/\beta^* + \delta$, where $\alpha$ and $\delta$ are fitting parameters.

In Fig. 6 one finds the dependence on reduced temperature of the fitting parameter $\alpha$, used to fit the dependence of the diffusion coefficient on the parameter $\beta^* = \omega_C^*/\omega_P^*$, where $\omega_C^*$ and $\omega_P^*$ are the cyclotron and plasma frequencies, respectively. The data and the solid blue line in Fig. 6 show that $D^* \propto T^*$, for $T^*/T_C^* > 1$ for weak coupling and for the values of the cyclotron frequency used in this work.

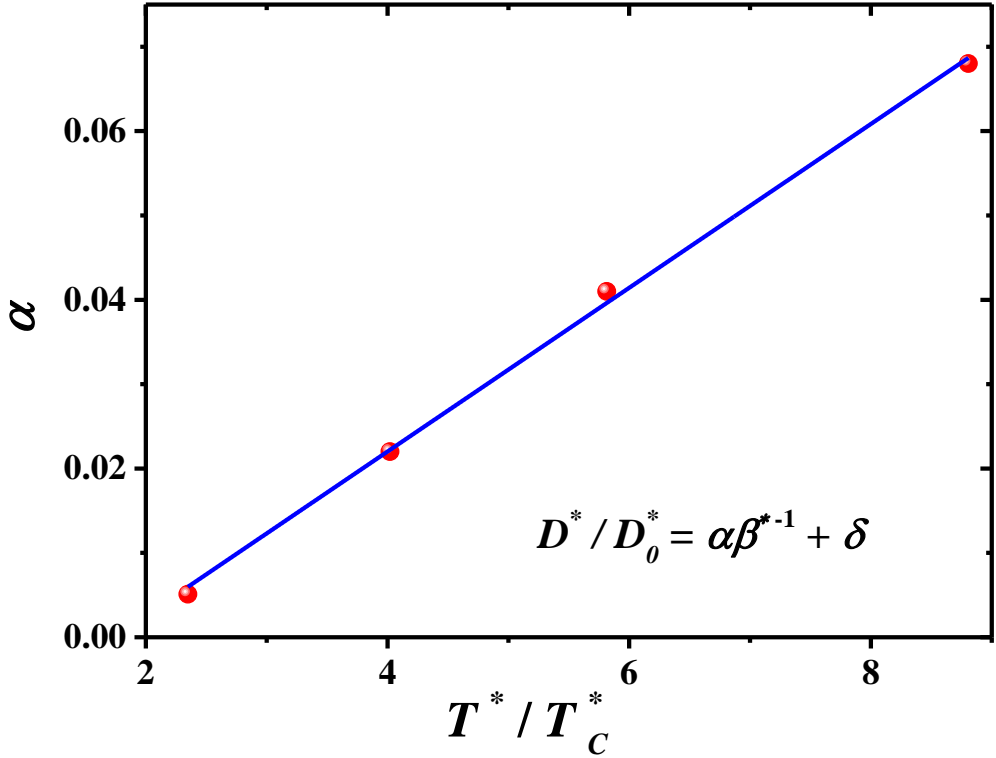

**Figure 6.** Temperature dependence of the parameter $\alpha$ used to fit the self-diffusion constant, $D^*$, to the equation $D^*/D_0^* = \alpha\beta^{*-1} + \delta$ (see also Fig. 5), where $\beta^* = \omega_C^*/\omega_P^*$, and $\delta$ is a fitting parameter. Here, $\omega_C^*$ and $\omega_P^*$ are the cyclotron and plasma frequencies, respectively. The solid blue line is the best linear fit.

Below $T_C^*$ all charges are condensed into a single structure with $q^* = 0$, thus there is no contribution to the dynamics from the Lorentz force. Therefore, we focus on the properties of $D^*$ for increasing magnetic field, at $T^* > T_C^*$. To extract the influence of the magnetic field and temperature on the average self-diffusion coefficient of the charged spheres, $D^*$ is plotted as a function of the coupling constant, $\Gamma^*$, for increasing $\beta^*$, in Fig. 7. Since the charge, mass and number density remain constant, increasing $\Gamma^*$ is equivalent to reducing the temperature, while larger $\beta^*$ corresponds to stronger magnetic field. The solid lines in Figs. 7(a)-7(f) are the best fits to the function

$$D^* = \frac{A}{\Gamma^*}e^{-b\Gamma^*} + c, \tag{2}$$

obtained by Daligault [35] for 3d one-component plasmas whose diffusion is driven by thermally activated jumps between equilibrium configurations ("cages") separated by energy barriers. *A*, *b* and *c* in eq. (2) represent adjustable parameters. Lowering the temperature and increasing the magnetic field (higher values of $\Gamma^*$ and $\beta^*$, respectively) results in a pronounced decrease in the self-diffusion coefficient. The fits to eq. (2) in Fig. 7 show that at high temperature (low $\Gamma^*$), $D^* \sim (Aa^*/q^{*2})T^*$, which agrees with the temperature dependence found for the self-diffusion coefficient of strictly 2d, unmagnetized charges [32,36].

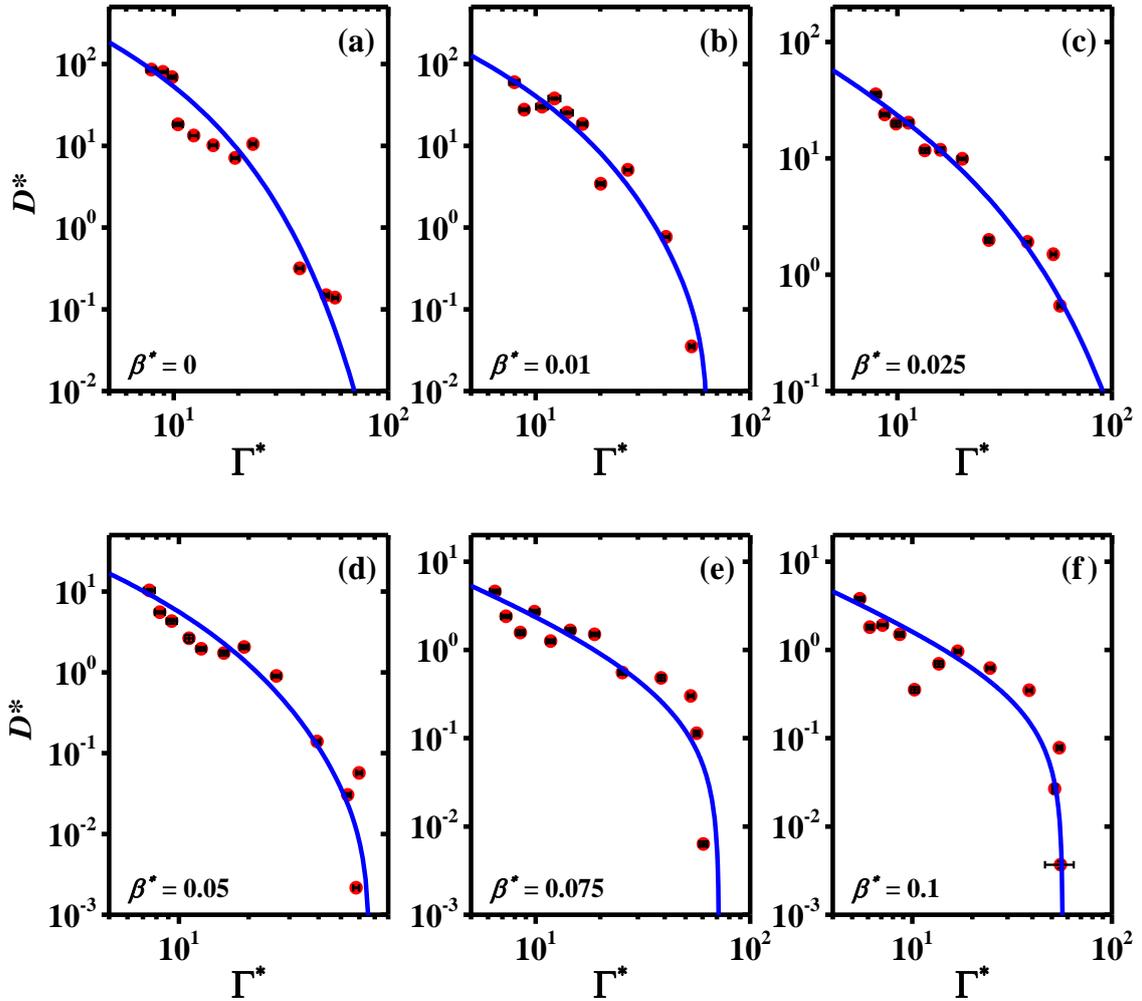

**Figure 7.** The average self-diffusion coefficient of all the charged spheres, $D^*$, as a function of the coupling constant, $\Gamma^*$, for increasing values of the parameter $\beta^* = \omega_C^*/\omega_P^*$,

for $T^*/T_C^* > 1$ in all cases. The data from the simulations are shown in solid (red) circles. The solid (blue) lines are the best fit to eq. (2). Here, $\omega_C^*$ and $\omega_P^*$ are the cyclotron and plasma frequencies, respectively.

## IV CONCLUSIONS

Consistent with the BvL theorem for static and thermodynamic properties, our numerical results indicate that a transverse magnetic field applied to a quasi-2d neutral set of charged spheres influences mostly their diffusion, with the structural properties turning out to be almost insensitive to the field. In particular, the critical or melting temperature is unaffected by the presence of the magnetic field, as is the melting process itself, for low density systems. The spatial correlations below and above the critical temperature under the applied magnetic field conserve their behavior as it was without field. The rate of dipole-pair breaking under a transverse magnetic field with temperature follows the same trend as that found without magnetic field. The BvL theorem does not apply to dynamic properties which we explore here for the first time. These exhibit marked changes as a result of the application of the external field. The self-diffusion coefficient is strongly influenced by the magnetic field. As the magnitude of the magnetic field grows, the self-diffusion coefficient decays as the inverse of the field (Bohm diffusion) for relatively weak coupling, in agreement with experiments on dusty plasmas. To the best of our knowledge, this is the first work that explores the influence of magnetic field on the structural and dynamical properties of a low density, quasi-2d topological phase transition in a Coulomb gas and should afford a better comparison with experiments, which are never carried out in strictly 2d.

## V ACKNOWLEDGEMENTS

This project was sponsored by CONAHCYT through grant number 320197. J.D.H.V. also thanks CONAHCYT for a postdoctoral scholarship. Z. N. was partially supported by a Leverhulme Trust International Professorship grant (No. LIP-2020-014).

# Supplemental Material

# for

# Self-Diffusion and Structure of a Quasi Two-Dimensional, Classical Coulomb Gas Under Increasing Magnetic Field and Temperature

J. D. Hernández Velázquez[1], Z. Nussinov[2,3], and A. Gama Goicochea[1*]

[1]Tecnológico Nacional de México, Tecnológico de Estudios Superiores de Ecatepec, División de Ingeniería Química y Bioquímica, Ecatepec de Morelos, Estado de México 55210, Mexico

[2]Department of Physics, Washington University, St. Louis, Missouri 63130, USA

[3]Rudolf Peierls Centre for Theoretical Physics, University of Oxford, Oxford OX1 3PU, United Kingdom

## Abstract

Here, full details are provided about the dissipative particle dynamics force field used to solve the motion of the charged spheres. Numerical simulation details are also provided, as well as snapshots of the system and some additional results.

---

[*] Corresponding author. Email: agama@alumni.stanford.edu

# I. Dissipative particle dynamics

In the DPD model, the total force of each particle is the sum of three fundamental forces, a conservative ($\mathbf{F}_{ij}^C$), a dissipative ($\mathbf{F}_{ij}^D$) and a random force ($\mathbf{F}_{ij}^R$):

$$\mathbf{F}_{ij} = \sum_{i \neq j}^{N} [\mathbf{F}_{ij}^C + \mathbf{F}_{ij}^D + \mathbf{F}_{ij}^R]. \tag{S1}$$

The conservative force is given by a soft, linearly decaying repulsive function:

$$\mathbf{F}_{ij}^C = \begin{cases} a_{ij}(1 - r_{ij})\hat{\mathbf{r}}_{ij} & r_{ij} \leq r_c^* \\ 0 & r_{ij} > r_c^* \end{cases}, \tag{S2}$$

let $\mathbf{r}_{ij} = \mathbf{r}_i - \mathbf{r}_j$, $r_{ij} = |\mathbf{r}_{ij}|$, $\hat{\mathbf{r}}_{ij} = \mathbf{r}_{ij}/r_{ij}$, where $\mathbf{r}_{ij}$ the relative position vector between the $i$-th and $j$-th particles, and the parameter $a_{ij}$ is the maximum repulsion strength between the $i$-th and $j$-th particles. The dissipative and the random forces are, respectively:

$$\mathbf{F}_{ij}^D = -\gamma \omega^D(r_{ij})[\mathbf{r}_{ij} \cdot \mathbf{v}_{ij}]\hat{\mathbf{r}}_{ij} \tag{S3}$$

$$\mathbf{F}_{ij}^R = \sigma \omega^R(r_{ij})\xi_{ij}\hat{\mathbf{r}}_{ij}, \tag{S4}$$

where $\sigma$ and $\gamma$, are the noise amplitude and the friction coefficient indicating the maximum strength of the dissipative and random force, respectively. These two parameters are related through the relationship: $k_B T = \sigma^2/2\gamma$; where $k_B$ is Boltzmann's constant and $T$ the absolute temperature. In eq. (S3), $\mathbf{v}_{ij} = \mathbf{v}_i - \mathbf{v}_j$ is the relative velocity vector between the $i$-th and $j$-th particles, whereas in eq. (S4) $\xi_{ij} = \xi_{ji}$ is a random number uniformly distributed between 0 and 1 with unit variance. The weight functions $\omega^D(r_{ij})$ and $\omega^R(r_{ij})$ depend on distance and vanish for $r_{ij} > r_c$, and are chosen for computational convenience to be (see [1,2]):

$$\omega^D(r_{ij}) = [\omega^R(r_{ij})]^2 = max\left\{\left(1 - \frac{r_{ij}}{r_c^*}\right)^2, 0\right\}, \tag{S5}$$

where $r_c$ the cutoff radius of the three forces (eqs. (S2)-(S4)).

In the DPD model, $r_c^*$ is the inherent length scale and it is regularly chosen as the reduced unit of length, $r_c^* = 1$. Additionally, the values of the parameters of the dissipative and the random forces are chosen as $\gamma = 4.5$ and $\sigma = 3$, respectively; so that the internal temperature of the system is $T^* = 1$, this latter is used as the reduced unit of energy in DPD model [3].

To achieve the high degree of confinement at which all systems were subjected to obtain a quasi-two-dimensional regime, the systems were flanked by two implicit square surfaces placed at the ends of the *z*-direction of the simulation box. The force exerted over all particles by the effective walls is represented by a unidirectional, linearly decaying force law [4], in much the same spirit as the DPD conservative force and is shown in eq. (S2):

$$\mathbf{F}_{wall}(z_{iw}) = \begin{cases} a_w(1 - z_i/z_c^*)\hat{\mathbf{z}} & z_{iw} \leq z_c^* \\ 0 & z_{iw} > z_c^*, \end{cases} \tag{S6}$$

where $a_w = 120.0$ is the maximum repulsive strength exerted by any of the two walls over each DPD particle, which depends on the separation distance between the *i*-th particle and the wall, $z_{iw}$, and the cutoff distance, $z_c^* = 1\, r_c^*$; $\hat{\mathbf{z}}$ is the unit vector along the *z*-axis. This soft, repulsive force model has been implemented successfully in DPD simulations to study different phenomena under confinement [4–7].

The electrostatic interaction between particles is modeled by the implementation of the Ewald sums method [8,9] using charge distributions instead of point charges [10]. The charge distributions are defined as a Slater-type, radially decaying charge density function, given by

$$\rho_{q^*}(r^*) = \frac{q^*}{\pi\lambda^3} e^{-2r^*/\lambda}, \tag{S7}$$

where $\lambda$ is the decay length of the charge [10,11]. When the charge density, $\rho_{q^*}$ in eq. (S7) is integrated over space one finds that the total charge in the particle is $q^*$. In general, the forces between charge distributions cannot be calculated analytically. However, for the model given by eq. (S7) there is an accurate approximate expression that has been successfully tested in various applications [11–15]. The magnitude of the electrostatic potential between two charge distributions is defined by [10,11]:

$$U(r^*) = \frac{\Gamma}{4\pi} \left(\frac{q_i^* q_j^*}{r^*}\right) \left[1 - (1 + \kappa r^*) e^{-2\kappa r^*}\right], \tag{S8}$$

while the force derived from it is given by

$$\mathbf{F}_{ij}^E(r^*) = \frac{\Gamma}{4\pi} \left(\frac{q_i^* q_j^*}{r^{*2}}\right) \left[1 - (1 + 2\kappa r\{1 + \kappa r^*\}) e^{-2\kappa r^*}\right] \hat{\mathbf{r}}_{ij} \tag{S9}$$

where, $\Gamma = e^2/k_B T \varepsilon_0 \varepsilon_T r_c$ and $\kappa = r_c/\lambda$; $e$ is the electron charge, and the constants $\varepsilon_0$ and $\varepsilon_T$ are the dielectric constants of vacuum and water at room temperature, respectively. When Ewald sums are applied to confined systems, as is our case, additional care must be taken because the Fourier transforms involved cannot be performed straightforwardly due to the lack of three-dimensional periodicity [16]. This problem can be overcome by applying an additional force along the $z$-axis to all the charged particles:

$$\mathbf{F}_i^M(z) = -\frac{\Gamma}{V} q_i^* M_z \hat{\mathbf{z}}, \tag{S10}$$

where

$$M_z = \sum_{i=1}^{N} q_i^* z_i. \tag{S11}$$

In eq. (S10) $V$ is the volume of the simulation cell and $M_z$ is the total dipole moment, see eq. (S11), which must be removed out of the simulation cell for each particle. If the force in eq. (S10) is applied to each DPD particle [12,16], one can successfully implement the three-dimensional Ewald sums to confined DPD particles with charge distributions. Full details can be found in the work of Alarcón *et al*. [12].

The contribution of the external magnetic field applied in the *z*-direction of the simulation box, $\mathbf{B}^* = B_Z^* \hat{\mathbf{z}}$, where the Lorentz force has its usual form:

$$\mathbf{F}_i^B = q_i^* (\mathbf{v}_i \times \mathbf{B}^*) \tag{S12}$$

where, the magnitude of the magnetic field, $B_Z^*$, goes from $B_Z^* = 0$ to $B_Z^* = 0.1$.

## II. Simulation details

All the simulations reported here are performed in reduced units and under canonical ensemble conditions ($NVT$). The total number of particles, $N$, in the systems modeled is $2 \times 10^2$, except for the calculation of the radial distribution functions, where $3 \times 10^4$ DPD particles were used in the simulation box. The volume of the simulation box is $V = L_x^* \times L_y^* \times L_z^* = 80 \times 80 \times 1 \, r_c^{*3}$, for $N = 2 \times 10^2$, and $V = 10^3 \times 10^3 \times 1 \, r_c^{*3}$, for $N = 3 \times 10^4$. Periodic boundary conditions are applied along the *x*- and *y*-directions of the simulation box but not along the *z*-direction because the system is confined in that direction. To integrate the equations of motion we use a version of the Velocity-Verlet algorithm adapted to DPD model [17]. The time step chosen to integrate the equation of motion is $\Delta t^* = 0.03\tau$. All simulations were run for at least $1 \times 10^7$ time steps and up to $4 \times 10^7$ time steps, with the first half used to reach equilibrium and the second half used for the production phase. The simulations are carried out using a code developed by the group of one of us (A.G.G.), which has been benchmarked and thoroughly tested [12].

The real part of the Ewald sums is cut off at $r_E^* = 3.0\, r_c^*$, and $\alpha_0 = 0.15\, Å^{-1}$. The latter is the factor that determines the contribution in real space of the Ewald sums [10]. The value for $\lambda$ and $q^*$ in eq. (S7) are chosen in all cases as $\lambda = 6.95\, Å$ and $q^* = 4\, e$. For the part of the Coulomb interaction calculated in Fourier space, a maximum reciprocal vector $\vec{k}^{max} = (5,5,5)$ is chosen. The constants $\kappa$ and $\Gamma$ in eq. (S9) are chosen as $\kappa = 0.929$ and $\Gamma = 13.87$, following Terrón-Mejía *et al.* [10] and Groot [18], respectively. Because there are equal numbers of positively and negatively charged particles, the net charge in the system is always zero.

The self-diffusion coefficient, $D$, was calculated through the averaged mean-square displacements (MSD) of all DPD particles:

$$\text{MSD} = \left\langle \frac{1}{N}\sum_{i=1}^{N}|\mathbf{r}_i(t) - \mathbf{r}_i(0)|^2 \right\rangle = 6Dt, \tag{S13}$$

using the last $10^5$ time steps of each simulation to calculate the MSD.

The physical units of all quantities used in the simulations, have a conversion factor to SI units, which are related to the DPD units of mass ($m$), length ($r_c$), energy ($k_B T$), charge ($q$) and time ($\tau$). For instance, to attach physical units to the value of the magnetic field given in reduced units, the magnitude of the Lorentz force is used:

$$B_0 = \frac{F_0}{ev_0}, \tag{S14}$$

where the units of force are given by $F_0 = k_B T_0/r_c$. The units of velocity are $v_0 = r_c/\tau$. Using $T_0 = 1\, K$, as in typical relevant experiments [19,20], $r_c = 6.46\, Å$ [3] and the characteristic time $\tau = 3\, ps$, one finds that $B_0 = 619.26\, T$. Thus, the magnetic field used in our simulations is $B_Z = B_0 B_Z^*$; given the range of $B_Z^*$ values used in this work, we find that $0 \leq B_Z \leq 61.92\ T$. Table 1 displays the values of the main DPD units used in the current work, and their conversion factors to SI units.

**Table 1**. Table of the main reduced DPD units used in this work, and their conversion factors to SI units.

| | DPD unit | SI conversion factor |
|---|---|---|
| Mass, $m^* = 1$ | $[m]$ | $9 \times 10^{-23}$ g |
| Length, $r_c^* = 1$ | $[r_c]$ | $6.46 \times 10^{-10}$ m |
| Energy, $E^* = 1$ | $[k_B T]$ | $4.14 \times 10^{-21}$ J |
| Charge, $q^* = 4$ | $[e]$ | $1.602 \times 10^{-19}$ C |
| Time, $\Delta t^* = 0.03$ | $[\tau]$ | $3.012 \times 10^{-12}$ s |
| Magnetic field strength, $B_Z^* = 0.01, 0.025, 0.05, 0.075, 0.1$ | $[B_0]$ | 619.26 T |
| Self-diffusion coefficient, $D^*$ | $[r_c^2/\tau]$ | $1.392 \times 10^{-7}$ m$^2$/s |

The cyclotron and plasma frequencies are given by their usual definitions, namely $\omega_c^* = \frac{q^* B_Z^*}{m^*}$ and $\omega_p^{*2} = \frac{\rho^* q^{*2}}{\varepsilon_0 m^* a^*}$, respectively.

## III. Additional results

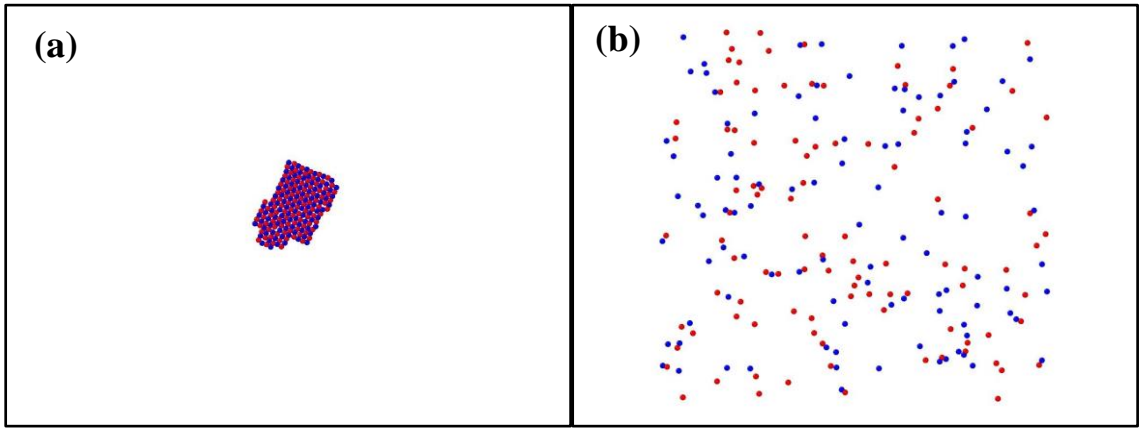

**Figure S1.** Snapshots on the *xy*-plane of the system of $N = 200$ charged spheres restricted to move under quasi-2d conditions (a) below $T_C^*$ and (b) above $T_C^*$ ($T^*/T_C^* = 0.75$ and $T^*/T_C^* = 3.75$, respectively). Red spheres represent positively charged particles, while blue spheres are negatively charged. For both cases, the magnetic field applied is $B_Z^* = 0.1$. The snapshots were obtained using OVITO [21].

In Fig. S1 one finds snapshots of the quasi-2d system of charged spheres (a) below, and (b) above the transition temperature, under the maximum strength of the magnetic field modeled in this work. All particles are condensed into a single structure below $T_C^*$, while most of them are unpaired above it. See also [13].

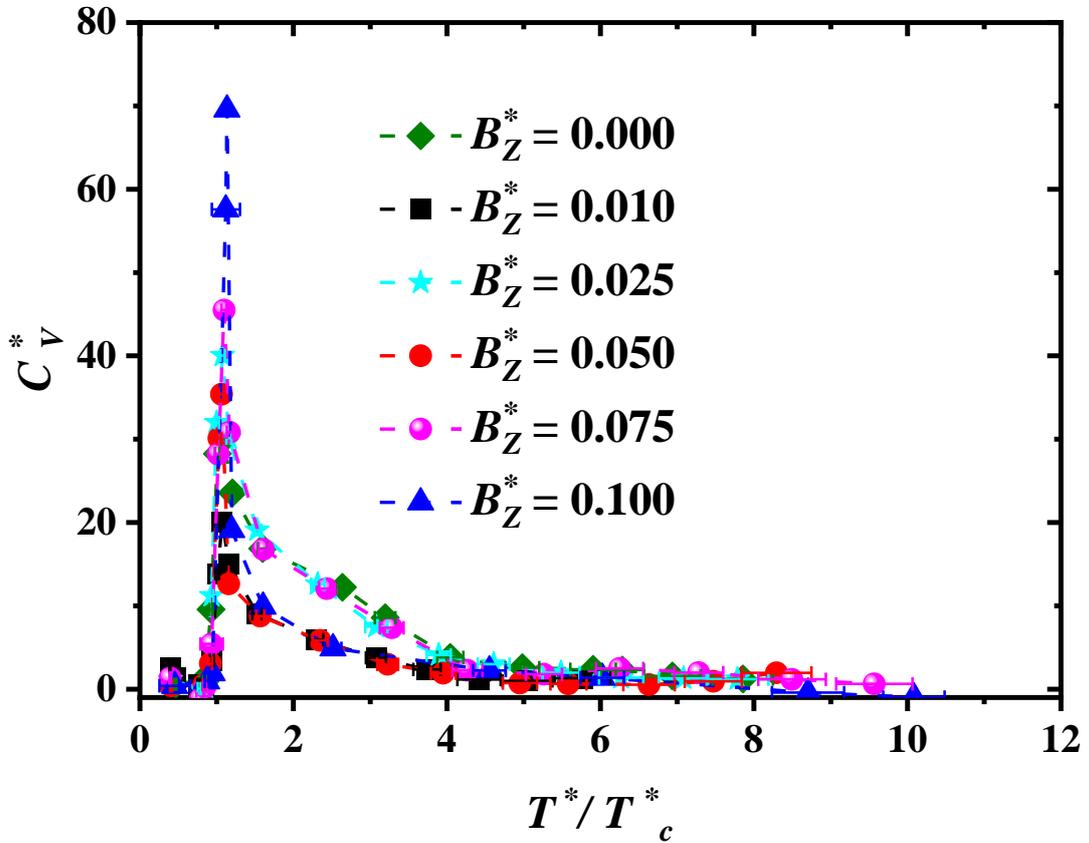

**Figure S2.** Specific heat dependence on temperature for the system, as the intensity of the magnetic field is increased. The temperature is normalized by $T_C^*$. The dashed lines are only guides for the eye. The data for $B_Z^* = 0.0$ were taken from [13].

The specific heat $C_V^*$ dependence of reduced temperature under increasing magnetic field is shown in Fig. S2. To obtain $C_V^*$, the total energy was calculated for all temperatures and its numerical derivative was acquired as differences in energy over differences in temperature, [13]. The application of the magnetic field does not change the melting temperature of the system, which always occurs slightly above $T_C^*$, as in the unmagnetized systems [13].

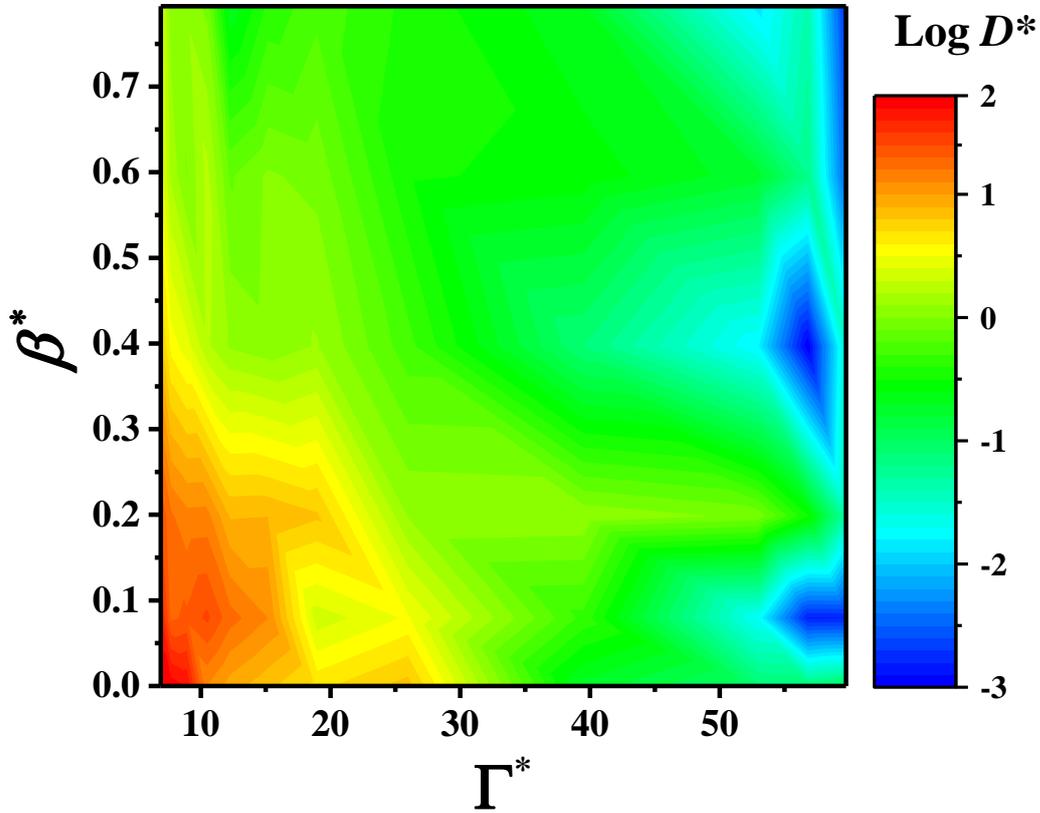

**Figure S3.** Map of the self-diffusion constant as a function of the coupling constant $\Gamma^*$ and the parameter $\beta^* = \omega_C^*/\omega_P^*$. The color bar on the right shows the range of values of the self-diffusion parameter in logarithmic scale, for clarity.

Lastly, in Fig. S3 there is a map showing how the self-diffusion coefficient changes as a function of the variables $\Gamma^*$ and $\beta^*$, simultaneously. The bar on the right of the figure shows the range of values of the self-diffusion coefficient, in logarithmic scale. The highest values of $D^*$ are obtained in the region of low $\beta^*$ and low $\Gamma^*$, i.e., for small values of the magnetic field and at high temperatures, respectively.